\documentclass[8pt,conference]{IEEEtran}
\usepackage[utf8]{inputenc}
\usepackage{subcaption}
\usepackage{subfig}
\usepackage{graphicx}
\usepackage{xcolor}
\usepackage{algorithm}
\usepackage{algorithmicx}
\usepackage{algpseudocode}
\begin{document}	
\title{WARP: A ICN architecture for social data}
\author{
\IEEEauthorblockN{Fabio Angius\IEEEauthorrefmark{1},  Cedric Westphal\IEEEauthorrefmark{2}\IEEEauthorrefmark{3},  Mario Gerla\IEEEauthorrefmark{1}, Giovanni Pau\IEEEauthorrefmark{1}\IEEEauthorrefmark{4}}\\
\IEEEauthorblockA{\small\IEEEauthorrefmark{1}Department of Computer Science, University of California, Los Angeles, California (USA) }
\IEEEauthorblockA{\small\IEEEauthorrefmark{2}Innovation Center, Huawei, Santa Clara (USA) }
\IEEEauthorblockA{\small\IEEEauthorrefmark{3}Department of Computer Engineering, University of California, Santa Cruz}
\IEEEauthorblockA{\small\IEEEauthorrefmark{4} Universit\`{e} Pierre et Marie Curie (UPMC) - LIP6,  Sorbonne Universites - Paris, France. \\ $<$fangius, gerla$>$@cs.ucla.edu, cedric.westphal@huawei.com, giovanni.pau@lip6.fr}
}
\maketitle
{\abstract Social network companies maintain complete visibility and ownership of the data they store. However users should be able to maintain full control over their content. For this purpose, we propose WARP, an architecture based upon Information-Centric Networking (ICN) designs, which expands the scope of the ICN architecture beyond media distribution, to provide data control in social networks. The benefit of our solution lies in the lightweight nature of the protocol and in its layered design. With WARP, data distribution and access policies are enforced on the user side. Data can still be replicated in an ICN fashion but we introduce control channels , named \textit{thread updates}, which ensures that the access to the data is always updated to the latest control policy. WARP decentralizes the social network but still offers APIs so that social network providers can build products and business models on top of WARP. Social applications run directly on the user's device and store their data on the user's \textit{butler} that takes care of encryption and distribution. Moreover, users can still rely on third parties to have high-availability without renouncing their privacy.
}

\section{Introduction\label{lbl:intro}}

Online social networks (OSNs) such as Facebook, Twitter, Google+, Myspace, LinkedIn, etc. have become irreplaceable for staying in touch with friends. The reason behind their success is that they enable forming virtual communities. As a consequence the data shared over an OSN can be very personal and demand immediacy from the provider.

OSNs gather huge amount of data about their users, and are now the real custodians of people's identity. Some OSNs even allow third party's software to use this knowledge\footnote{An example is the protocol Facebook Connect \cite{facebookapi}.} in order to improve the user experience. Social data can also be used to infer people's interests in order to win their attention and deliver messages more effectively, such as advertisements~\cite{socialcamp}. The OSN platform can also be used by governments and legal agencies to observe and monitor the user's activities \cite{nsa}.

Despite the wide success of social networking, there are controversial aspects to delegating the ownership of social content to the OSNs providers. Currently OSNs maintain ownership of both the social graph and the shared data~\cite{sdataown} and they often share them for profit or for legal interception, sometimes without the explicit consent of the user. Further, some information regarding one user might be subject to the privacy settings of one of his social connections, resulting in privacy leaks.

Recent research efforts \cite{prpl,tribler,junction,musubi}, as well as commercial initiatives \cite{tent}, have proposed to decentralize the social network so that users can maintain control over their data.
Concurrently, the research community has developed new ideas for an Internet architecture that can better serve the needs of users and developers. In particular, Information Centric Networking (ICN) architectures focus on connecting users to data instead of setting up connections between machines. Namely, ICN architectures aim at decoupling the data from a specific location in the network and at enabling location/identity separation: a file has a unique name that is independent from the address of the machine where it is stored, and which is used to route to any copy of the data. By doing this, it is possible to fetch the data from any of its replicas in the network and to cache data on any router in order to prevent unwarranted network traffic.

It is our belief that ICN architectures are now mature enough to overcome the network issues that distributed social networking has faced so far. Therefore, we propose a solution that implements all the functionality common to social networks while leaving to the users in full control of their data. Furthermore, our solution is compatible with the value add of current OSN providers.

In this paper we present the WARP framework, that is: an architecture that allows users to share data, potentially over the infrastructure of an OSN provider, but which prevents the OSN provider to access to the user's data without the explicit consent of the user.

Let us, for instance, consider Facebook. At the moment, the "Facebook application" is a software that runs on their servers in their data center, and the end-user chooses how to interface to it, either via a web-browser or a mobile application. In other words, data is generated at the source by the user, and then moved to Facebook. Then Facebook manages and distributes the data. We want to reverse this scheme in an ICN fashion by having a Facebook application running on the source of the data, i.e. the user device. The application must have the duty of collecting the data from the users, whatever this data is, and pass it to the WARP framework which then takes care of encrypting and distributing the data to the network as a authenticated source.

Note that Facebook can still distribute the data, for instance to improve the performance or offering the distribution service in exchange of showing their advertisement. WARP only guarantees to the user that his data will not be accessible unless he decides to grant to Facebook the right to access it. Our ultimate goal is to show that, with minor changes, ICN does this by design, and that additionally, it is possible to use WARP to build a shared social infrastructure as an overlay of the current Internet that is able to serve both the interests of OSNs and the rights of the users.


The rest of the paper is organized as follows: in the next section, we review the related work for both privacy preserving OSN mechanisms and Information-Centric Networks. In Section~\ref{sec:challenge}, we provide motivation and technical requirements for our architecture. We then describe our architecture in Section~\ref{sec:arch}, with the details of the protocol in Section~\ref{sec:protocol}. We evaluate the framework in Section~\ref{sec:eval}, before concluding the work in Section \ref{sec:end}.

\section{Related Work\label{sec:related}}
\subsection{Information Centric Networking}

Information Centric Networking\cite{icnsurvey} describes network architectures that use data retrieval primitives, i.e. \textit{put/get}, in place of the primitives for machine-to-machine message delivery, i.e. \textit{send/listen}. The final goal is to decouple applications from topology and fetch data from anywhere in the network, including transparent caches. Traffic in the Internet follows a power law distribution where a significant fraction of the traffic comes from a rather restricted number of items. For this reason, as the cost of storage decreases much faster than the cost of bandwidth, inserting storage within the network as proposed by ICN protocols appears to be a valid alternative to the current architecture.

ICNs usually rely on two types of packets, data and requests. Data packets simply contain the content object. The content is uniquely named, signed and, since there is no access control on the network caches, encrypted by their producer for security reasons. Requests, intuitively, are the messages used to fetch data. How to route content requests is a critical issue for ICN protocols. Some architectures, such as \cite{dona,netinf}, use name resolvers to locate content. This solution, however, is less responsive to changes and is exposed to attacks, such as DoS. Alternatively,  architectures such as \cite{van1, haggle} use routing tables or hybrid schemes. For name resolution, the reader is reminded of~\cite{vanglobally, vancus}.

By design, ICN architectures secure the data instead of the communication channel. However, since the producer has no control over the replicas of its data, their application to social networking is very limited. A trivial example is the following: let us imagine that a content, say A, is initially shared by Alice with all her friends. After a few of her friends have downloaded A, A is replicated on several distributors. If now Alice wants to withdraw the permission of reading A from Bob, she must first ensure that all the copies of A stored in caches are voided and replaced with the new version of A. This is required to avoid that Bob fetches the first version of A that he is able to decrypt. The task is made difficult for Alice since the data could have been replicated several times on caches that are unknown to her.

\subsection{Privacy Preserving Social Content Distribution}

PrPl \cite{prpl} and Musubi \cite{musubi}, have deeply inspired this work since, to the best of our knowledge, these were the first research projects that aimed at decentralizing modern social networks. Despite sharing the same goal, though, their nature is very different from WARP's. In fact, they focus on the application support, and specifically on distributed search indexing. WARP proposes a new network infrastructure for distributing data in an efficient and secure manner. For this reason WARP has to be considered complementary to these projects and integrating all solutions together is the final goal.

Another similar project is Tribler~\cite{tribler}. Tribler, differently from PrPl and Musubi, does take care of the connectivity aspects of distributed OSN although in a differently manner from WARP. It is not a generic architecture and cannot easily interface with the OSN software; moreover it relies on gossiping information across a P2P network and is intrinsically subject to high latencies and provides inadequate real-time notification support.

Cachet \cite{cachet} addresses a similar problem as WARP. It implements distributed social feeds, policy based encryption and even offline persistency (which WARP does not entirely support at the moment). However, Cachet has three substantial limitations: (1) by admission of its authors, the computational overhead needed to secure access and updates on the DHT is not sustainable; (2) it relies on proxies to re-encode the content and to revoke keys; and (3) it is limited to the news feed type of content whereas WARP allows to develop any kind of software and to enhance it with social functionality. For the sake of precision, Cachet boosts its performance by implementing a gossip-based social caching, although it cannot outperform WARP's structured cache search algorithm.

Other relevant projects include Diaspora \cite{diaspora}, PeerSon \cite{peerson}, Safebook \cite{safebook}, LotusNet \cite{lotusnet} and SCOPE \cite{scope}. These projects implements subsets of Cachet's functionality and for this reason will not be discussed further in this paper. The reader is referred to the original papers for the details.
	
\section{Architectural Requirements}
\label{sec:challenge}
\subsection{Requirements}

Aside from the business incentives issue of protecting the data from the big data analysis of the OSN platform, there are several technical issues in decentralizing the social network. Those corresponds to some requirements that our framework needs to satisfy. We outline these (in no order of importance):

$\bullet$ \textbf{Timing:} The Internet is today a real living network. Social content is produced as part of the daily routine and must be delivered in real-time to all its recipients. Real-time feeds, such as the \textit{facebook newsfeed}, usually carry small amounts of data including text, web-links, picture thumbnails, etc. However, they require almost immediate transfer of data, unlike, say, P2P content distribution mechanisms.\\
$\bullet$ \textbf{Visibility and privacy:} People share personal data daily - e.g. opinions, thoughts, political views, pictures, videos etc. -  with their friends and acquaintances but not necessarily with everyone. Current solutions rely on authenticated connections and Access Control List (ACL) \cite{acl} to discriminate whether a content can be downloaded or not. To implement the same method in a decentralized manner can be computationally unsustainable. However the ICN approach overcomes this issue by securing the data rather than the connection. We will expand upon this point in the rest of the paper.\\
$\bullet$ \textbf{Scalability:} Social networks sites commonly rely on proprietary large scale data centers for their storage. They also rely on Content Delivery Networks (CDN), such as Akamai, for content delivery. A sample architecture is represented. This solution is highly scalable and fault-tolerant but rarely leaves the user with full control over his data; the users cannot, for instance, decide where unencrypted data are stored, how many times it is replicated and, more importantly, how fast access to data can be revoked. All the previous works minimally use caching: each user is considered responsible for his own content and serves all the requests. In our context, each user is typically a recipient of several hundreds of feeds and also has as many followers. While the current framework uses the OSN servers as aggregators, a decentralized solution opens new scalability issues and a potentially unsustainable overhead. Our solution will have to satisfy this requirement.

\subsection{High level view of WARP}

We now describe a high level view of WARP. We propose an open infrastructure to decentralize the distribution of social data whilst enforcing user's privacy. WARP allows to disseminate social data in an ICN fashion, leaving to the user the opportunity of deciding the right tradeoff between performance, security and control on his own data. The protocol is based on a hierarchical name resolver to facilitate routing without lacking information about users' activities.

Data security is enforced using a combination of symmetric cryptography and Attribute Based Encryption (ABE) similarly to what is done in \cite{decent, easier} but with the only difference that WARP does not depend on proxies for content updates.

Content updates are taken care of by implementing a two-way agreement between the producer of the original content and the producer of the update. Namely, assuming the case of a comment on a Facebook post, the first user (who owns the post) will agree to link the content update to his Facebook wall as long as the second user gives a signed certificate of what its comment contains and what the name of the data is. By creating this link, the owner of the wall prevents two problems, first: it links only comments that it consider appropriate and reputable; second, it certifies to his audience that they have not been compromised afterwards; at the same time the content update is bound with its actual producer, who is also be responsible for its distribution.

Instead of implementing a key-revocation protocol, whenever security policies change, the cached objects are deleted, or substituted with a newly encrypted version. This is made possible by using control channels, named \textit{thread updates} which are used for delivering commands to the content distributors. Note that the thread updates are content and therefore are routed, requested and even cached (only by distributors).

It is important to remark that we enforce caching to take care of the bursty nature of content requests \cite{neworleans} more than for long term data distribution.

\subsection{A Motivating Example}

This section is meant to help the reader understand how WARP changes the normal workflow of social network. Let us consider the case of two users, Alice and Bob, who want to use a Facebook-like application (FL) built for the WARP architecture.

\begin{enumerate}
\item Both the users must have control of a {\em butler} for their online identity. The butler is the host of the user's social data as well as its guardian. This can be available both as paid service, ad-based service, or a always- connected set-top-box. For simplicity, hereafter we will consider this last case only, moreover even though WARP supports mobile devices, as described in section-, for the moment we will consider the content to be created on the butler as well.
\item Both users have to create their own WARP identity, registering to an Identity Authority (IA)that, at its own condition, does nothing more than sign and authenticate Alice's and Bob's public keys. The IA can be any institution that is widely considered reputable and trustworthy, including an OSN of the user's choice.
\item Without regard to what social software Bob and Alice are going to use they are already able to establish a social relationship as a mere exchange of keys between their butlers. Friendships are defined by assigning a user to one or more categories - e.g. "friend", "colleague" or "family". Moreover each user always belongs to a group consisting of himself only. Let us assume that Alice categorizes Bob as friend whereas Bob categorizes Alice as "colleague".
\item Alice and Bob install on their butler the FL software which has the following duties: offer a user interface, compile the data in the application format, interface with the butler's database to store the data thereof, choose what groups can access to the content created.
\item Alice has not privacy issues; however she does not want to use her own bandwidth to distribute her FL data. Therefore she chooses to distribute everything named under {\em Alice/facebook/} using the FL infrastructure (the service can be offered behind payment or as ad-supported).
\item Bob is not concerned about bandwidth consumption and does not want to pay cloud service to distribute his data. Instead he organizes a P2P network with his closest friends that become distributors of his content. in exchange he does the same. (IMPORTANT: Links are chosen based on well-known literature on tit-for-tat, social routing, social caching etc.)
\item Alice shares her newsfeed with all her friends and so does Bob. In this case, Bob will be able to follow Alice's feed whereas Alice cannot do the same with Bob's, moreover Alice's will not even be able to see if Bob is writing on his wall because she will not be able to identify what the data objects are.
\item When Bob wants to see Alice's stream the FL application takes care of identifying all the chunks of the stream, download them, decrypt them and print them on the screen aggregated with all news of Bob's friends.
\item Whenever Alice wants to publish something that Bob should not see she can add to the feed a post with a content specific policy, that excludes Bob from the audience. For instance by replacing the group "friend" with the id of every single friend beside Bob, or creating a new group and distributing the new keys to its members. In this case Bob will be able to see that the network object (because he has access to the stream). However he will not be able to decrypt the post.
\item In order to change the access to a content that has already been published, the user has nothing else to do than simply change the policy on it. The butler will take care of voiding every cached copy. The new copy will be distributed by the butler upon request. Note that when a member is removed from a group, the keys for that group must be redistributed in order to avoid a massive key exchange. For this scenario, we propose a load balancing solution, see section \ref{sec:keyrevoc}.
\end{enumerate}

As we can see in this example, both users select the privacy level they would like to achieve. Users are able to change privacy setting, and to revoke authorization to access content, even though this content has been distributed in the wild.

\section{Architecture\label{sec:arch}}
\begin{figure*}[t!]
\begin{minipage}[b]{0.49\linewidth}
\centering
\includegraphics[width=\linewidth, height=4cm]{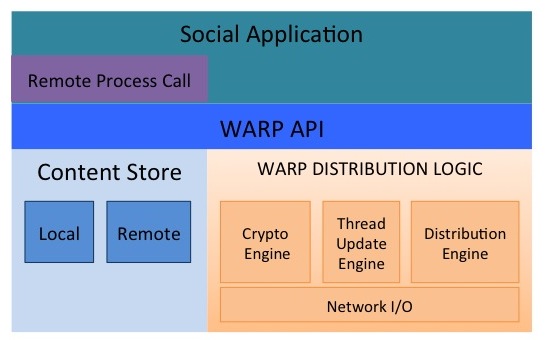}
\subcaption{\scriptsize{A sketch of the internals of a butler. Note that the social application can interact locally or remotely via RPC.}}
\label{fig:arch}
\end{minipage}
\begin{minipage}[b]{0.49\linewidth}
\centering
\includegraphics[width=0.8\linewidth, height=4cm]{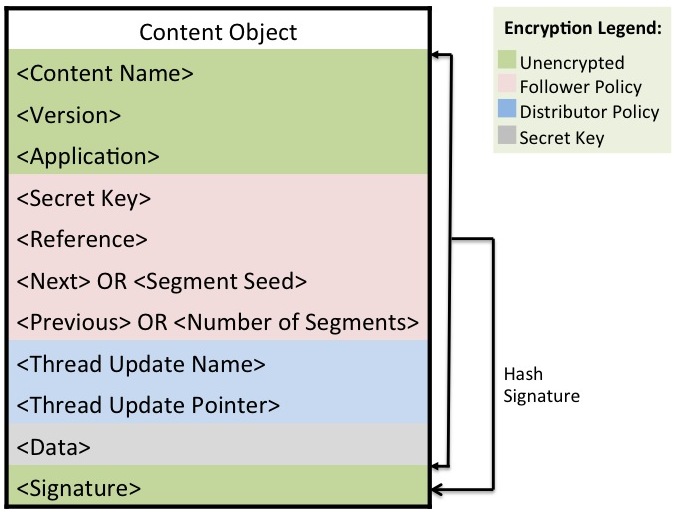}
\subcaption{\scriptsize{The structure of a network object}}
\label{fig:object}
\end{minipage}
\caption{{ }}
\end{figure*}

\subsection{Data Naming}

WARP organizes content in \textit{network objects} that are uniquely named using the following structure:
\begin{verbatim}
x.y/<application>/<folder>/.../<appendix>
\end{verbatim}
There must be at least three segments divided by a backslash. The first one is always in the form $x.y$ where $y$ is the user name of the producer and $x$ is the IA that certifies his identity and signs his public keys. The segment application is a unique identifier for the software that compiled the content. The remaining part of the name is organized as file-system, the segments in the middle are \textit{folders} and are used for both organization and routing purposes. As last, the \textit{appendix}, identifies the network object within its folder. Segments are named using a random generator to make content untraceable and avoid leaking information about its nature, for instance a picture named "Me and Paul at Hawaii", only the username and the application name is kept in a human readable format.
\subsection{Butlers and Distributors}

A WARP network is a two-tier overlay composed of \textbf{\textit{butlers}} and \textbf{\textit{distributors}}. 

As in~\cite{prpl}, a butler is a service that can be run on a set-top-box at home, or as a paid or ad-supported service by a vendor of choice. Every user in the social network must have a \textbf{butler} that serves four fundamental tasks: establishing social links with the other butlers, interfacing with the social applications; storing the user's data; and organizing data distribution.

Users establish their social relationship directly from the butler interface. The OSN accounts will be dismissed, so that there will be only one global social graph. A relationship is established when the user categorize the identity of another user. Thanks to the expressivity of ABE~\cite{abe}, this allows both symmetrical, as in Facebook, or asymmetrical, as in Twitter, relationship.
A sketch of a butler's internals is showed in Fig. \ref{fig:arch}. At the top of the stack, there is the social application that interfaces with the butler locally or via Remote Process Call. The latter is a strong requirement of our design since content is often generated and consumed from mobile devices.
WARP APIs only allow the social application to store their data in the local storage and communicate whether they should be made available to the network. The data distribution and encryption are entirely managed by the WARP DISTRIBUTION LOGIC. The social application cannot decide, for example, who will be able to see the data and not even what the data name is going to be. 

A butler is always the root source of the user's content. It can distribute the data as a stand-alone server or using \textit{distributors} that behave similarly to ICN caches. A distributor can be: any trustworthy butler, if users want to share their data in a pure peer-to-peer fashion; a paid cloud service, if performance are a concern; or an ad-based service offered, for instance, by the company that developed the social application. In order to be become a distributor a host receives a Distribution Certificate:
$$Dist_C=\{x.y, f, D_{f}, D_K, T_{exp}\}_{P_{K_{x.y}}}$$
where $x.y$ is the butler identity, $f$ is the folder that must be distributed, $D_f$ is a list of all the distributors of $f$ and all its subfolders, $D_K$ is a key that must be used to decrypt part of the content that are for the distributor's view only, and $T_{exp}$ is the expiration date of the certificate--the certificate is signed with the public key of the user. A key point of our architecture is that a distributor is not necessarily a single machine, a distributor can be the entire Akamai network for instance. An IA must certify the distributor identity and which machines are attributable to it. WARP does not specify how a distributor must balance the load of request across its servers.

WARP's distributors differ from ICN caches because they must monitor the updates of the content they store via control channels, named \textit{thread update}, that deliver information about the cached content. Thread updates deliver two types of information: (1) notify security updates, e.g. a newly encrypted version of the content, and (2) notify the creation of new related content in order to allow prefetching.

\subsection{Cryptography and Credentials}

Content is authenticated with signatures generated using any public-key scheme, such as DSA \cite{dsa} or ECDSA \cite{ecdsa}, that guarantees confidentiality\footnote{not all the schemes guarantee confidentiality. For instance, Identity Based Encryption (IBE) \cite{ibe} does not}. At the current state of the art we rely on the IA to authenticate the public keys. However we conjecture that, given the nature of WARP, PGP \cite{pgp} and a web-of-trust \cite{weboftrust} can suffice for the task. In any case we consider the problem of key authentication beyond the scope of this paper.

Privacy, is enforced using a combination symmetric and attribute based encryption (ABE) \cite{abe}. ABE is an asymmetric scheme adapted for broadcast encryption \cite{broadcastenc}. It gives fine control over the audience of a private message. The cyphertext is encrypted with a \textit{policy} expressed as a boolean expression. Variables, named \textit{attributes}, are connected by the operators $AND (\wedge)$, $OR (\vee)$ and \textit{k-of-n}. Every user possesses a private-key that is associated with some of such attributes and is able to decrypt the cyphertext if these satisfy the encrypting policy. For example, let us consider a secret $s$ that should be read only by friends, school acquaintances and teammates, $s$ will be encrypted with a policy: $$P=friend\vee(acquaintance\wedge school)\vee teammate$$.
$P$ allows anybody with the attribute "friend" or "teammate" to decrypt, although the attribute "acquaintance" alone will not suffice unless it is combined with "school".\\
Performance-wise, ABE, as most of asymmetric encryption schemes, is sensibly slower than symmetric cryptography thus WARP secures the data using a secret key $(sk)$ that is encrypted with ABE and distributed together with the cyphertext.

\subsection{Key-distribution}

Butlers exchange keys on a as-needed basis, only when some content has to be decrypted and not when the users define their relationship. The reason for doing so is that, as in \cite{secureattributebased}, keys have an expiration date, in the order of a week or a month. If two users do not interact often it is pointless to exchange keys. For the same reason, users should maintain a pool of recent keys in the eventuality that they have to decrypt a content that was encrypted with one of the previous keys. The key point here is that WARP does not revoke keys in the proper sense but instead revokes the cyphertext from the caches.\\
The butler automatically assigns a random \textit{alias} to every known user so that content can be encrypted for a single user only--and also this because it is needed by our key-revocation scheme.\\
Each content stored in the butler has two  associate policies: a Follower policy (FP), that specify who can access the content and Distributor Policies (DP) that is used to encrypt the fields that are supposed to be read by the distributors only. Whether a distributor can also decrypt the content or not is up to the user.

\subsection{Key-Revocation\label{sec:keyrevoc}}

Key revocation at the scale of a social network is a critical matter due to the fact that users have often hundreds to a thousand friends \cite{persona}.
Without loss of generality, let us consider the case of a content $c$ encrypted using a single-attribute policy $p=friend$. In order to oust a user from the audience its producer must: (1) create a new version of the attribute, say $friend'$, (2) for each key containing the attribute $friend$ distribute a new one containing the new attribute $friend'$, (3) the butler re-encrypts $c$ with policy $p'=friend'$. Similarly to \cite{junod}, we work around this problem by using aliases of users' identities. \\
The butler assigns an extra attribute, named \textit{\textbf{bucket}}, to each known user. A bucket is an integer between $0$ and $K$ where K is a constant parameter. When a user, say $B$, is revoked of the attribute, say $a$, the new policy will be such that the presence of $a$ will be conditioned to the presence of every bucket beside B's or the individual identity of every user included in B's bucket except B. This solution prevents the number of attributes from growing linearly with the number of users as long as the parameter K is chosen adequately. Moreover, since keys are redistributed periodically, the impaired policy will be enforced only until the new attributes will be generated.

\subsection{Content}
\begin{figure*}[t!]
\begin{minipage}[b]{\linewidth}
\centering
\includegraphics[width=\linewidth, height=4cm]{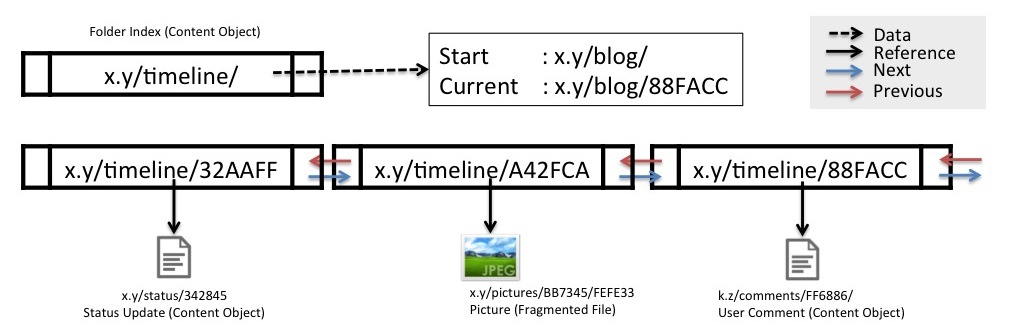}
\subcaption{\scriptsize{A possible organization of network objects to recreate a Facebook-like timeline. The index of the folder \texttt{x.y/timeline} keeps track of the latest segment of the news feed which is organized as a WARP feed. Each entry of the timeline reference an external object, from left to right: a status update from the user, a picture from another application, and the comment of another user. The social application "timeline" will take care of aggregating the content and show it to the user.}  }
\label{fig:appcon}
\end{minipage}
\end{figure*}

Data in the network is organized as a collection of \textit{network objects}, as in Fig. \ref{fig:object}, that can contain data, a reference to another object or both. Public fields contain general information about the object, they are: \textbf{Content Name},\textbf{Version}, \textbf{Application} and \textbf{Signature}. Followers fields are accessible only to followers, the most important is one \textbf{Secret Key} that is used to encrypt the data. \textbf{Reference} is used to redirect to another object while the other fields readable to the followers are use to link objects together and htey are explained in the next section. 

There are onyl two Distributors fields: \textbf{Thread Update Name}  and \textbf{Thread Update Pointer}. These are the name of the control channel for the content and a pointer to its latest message at the time the object was created. 

\subsubsection{Feeds, Fragmented Files and Indexes}

WARP essentially offers two data structures to organize content: Feeds and Fragmented Files.
A feed is a chain of objects doubly-linked with each other using the fields Next and Previous, as in Fig \ref{fig:appcon}. It is meant to be used for unbounded streams of data, for example the Facebook timeline. Each chunk of the stream is anonymized as any other name in the network and this is the reason why objects must be linked to each other. \\
Fragmented files are meant for bulky static data - such as pictures, audio, video etc - that are too big to be kept in one single content object and can be more efficiently encrypted and fetched as fragments. Every fragment has to be in the same folder, but, instead of being linked as a feed, their name can be generated using the \texttt{Segment Seed} contained within each fragment. For example, given a folder \texttt{/x.y/photos/AFAFA/)}, a seed $s$ and a hashing function $h$ the fragments name will be \texttt{/x.y/photos/AFAFA/h(seed,1))}, \texttt{/x.y/photos/AFAFA/h(seed,2))}; the number of fragments is known from the field \texttt{Number of Segment}.

This method is preferable to a feed because it allows fragments to be downloaded in parallel, although it has the disadvantage that if one fragments name is changed then also all the others must be changed (a feed can be modified more efficiently).

Up to this point we only discussed about objects that are anonymized, namely their name cannot be known before hand, although a social application needs to know what objects represent what, and what their name is. For this reason, WARP allows every application prefix (which are the only content names that are not anonymized) to answer with an object, named index, that work as an entry point for the application data. The object, can for instance maintain a pointer to the latest object of a feed, or maintain a directory of all the pictures stored in one's photobook.

\section{Protocol\label{sec:protocol}}
\subsection{Protocol Messages}

As discussed in Section \ref{sec:related}, ICN protocols generally use two types of messages, requests and data. WARP introduces two additional types of messages: \textbf{\textit{RESOLVE}} and \textbf{\textit{NOTIFY}}.

A \textbf{RESOLVE} message is used to obtain a list of distributors for a given folder. This type of query can be answered authoritatively only by the producer of the content or a distributor for a parent folder. In other words, considering the name \texttt{x.y/folder1/folder2/name1}
a butler that does not know a distributor for \texttt{x.y/folder1/folder2/} can obtain a list of distributors by sending a query to the butler \texttt{x.y} or any distributor for \texttt{x.y/folder1/}. This far we have not mentioned how to resolve machine addresses. Generally speaking since both butlers and distributors rely on the IA to certify their identity, as long as the identity is associated with a domain name the name can be resolved using the normal DNS infrastructure.

\textbf{NOTIFY} messages are one of the core ideas of WARP. They are sent by a butler to another to inform that a content has been created. They are fundamental to implement common functionality - such as comments, re-post, likes and re-tweets - since in WARP a network object cannot be modified by anybody that is not its producer. By design, content aggregation is operated at the application level rather than at the storage level. Namely, assuming the case of Alice adding a post on her Facebook wall and Bob commenting on it, in WARP there would be three distinct objects. Alice's post, Bob's comment and Alice's link to Bob's comment. A notify message contains a content-name, a signed checksum of the related data and an application identifier that says which application has generated the content.

\subsection{Thread Updates}

Thread Updates (TUs) are implemented as any other feed in the WARP network with the only difference that they deliver commands to caches.
Every network object is associated to a TU and carries a pointer to the latest object of the feed at the time it was created. Ideally content that are in the same TU should belong together so that for instance a comment belongs to the thread of the original post. However, we did not want to leave this kind of freedom to the developer and therefore we are using the following convention. Each folder is monitored by its own TU. However, since TUs are also content and can be monitored in the same way, each TU is monitored by the TU of its parent folder. This allows the distributors to have a finer control of what TUs to follow, for instance if a cloud service is distributing the entirety of a user's data it can follow a single thread for the user. If it is only distributing his photo album, it will monitor only the folder containing its pictures.

The commands that a distributor receives from a TU are the following:\\
\textbf{ADD(X)} X has been created. The distributor can decide whether to prefetch X or not. \\
\textbf{DELETE(X)} X must be purged from the cache and never distributed again.\\
\textbf{UPDATE(X,Y)} X has been updated with Y. The distributor must purge X, and can decide whether to download Y or not.\\
\textbf{CUT(X,Y,X',Y')} A feed has been cut from X to Y. All the objects in between must be purged from the cache, the distributor can decide whether to download X' and Y' to merge the remaining pieces.

For our future work we are considering implementing a command that distributes only a differential of an updated object containing renewed version of the secret key and a new signature. However there are security issues to take into account and we do not consider this optimization at the moment.

\subsection{Fetching of data}

Butlers and distributors maintain a routing table to direct their request. Each entry of the routing table contains a folder, a distributor and the expiration of the content. New requests that do not match any entry of the table trigger the butler to resolve the name and add a new entry to the routing table.

\subsection{Serving Data}
In the generic ICN architecture, whenever a cache, or a router, is requested for a content it verifies if there is a copy in its storage. Otherwise it forwards the request to another cache to obtain a copy. Whenever the copy arrives, it can decide whether it should be cached or not. WARP approach is slightly different since it must consider the thread updates. The logic for serving an incoming request for a content $R$:\\

\noindent\textbf{1.} The distributor verifies whether the content is in the Content Store, if not it continues at point 5.\\
\textbf{2.} Consider $TU_R$ the Thread Update of R, if $TU_R$ has been recently read the content is considered updated, then continue at 6. \\
\textbf{3.} Fetch $TU_R$. \\
\textbf{4.} Read $TU_R$ and update the Content Store accordingly, if the content is still updated continue at 7.\\
\textbf{5.} Fetch R from another distributor \\
\textbf{6.} Serve Request \\
\textbf{7.} If R is not a popular content then terminate.\\
\textbf{8.} If $TU_R$ or any of its parent TUs are balready being followed, add it to the list of threads that must be followed \\
\textbf{10.} Cache R in the Content Store \\

\subsection{Continuous Polling}

A possible attack that can tamper the cache control mechanism of WARP is \textbf{continuous polling}: a malicious butler generates requests at a high frequency so that it can receive every version of a content before being excluded from the followers.
This problem can be solved on the distributor side by implementing a \textit{Request Ban Window} (RBW), for every open connection distributors keep track of when each a butler is allowed to send a new request, if a request arrives before then they do not serve it.
Also, whenever they receive a request from the butler they update the RBW to the maximum between its current value and the value of a function:
$$T_{RBW}(t) = \mbox{current time} + \left\{
\frac{fail(t,\delta)}{1 + succ(t,\delta)}\cdot \delta \\
\right.$$
where $fail(t,delta)$ and $succ(t,\delta)$ respectively are the number of failed and successful requests that a butler has sent in the latest time window of length $delta$. Note that $T_{RBW}(t)$ scales the RBW $\delta$ according to the \textit{hit-ratio} of the butlers request, if the requests are mostly successful - i.e. there is a burst of data to download - the butler is allowed to send requests faster, differently, if the requests are unsuccessful the time window grows.
Butlers that respect the suggested inter-request time will never have their requests rejected whereas butlers that do not will delay their opportunity of downloading.

\subsection{Application}

Due to space constraints, we cannot give an extensive discussion on WARP's APIs although in Fig \ref{fig:archcon} we outline the content design of Facebook-like timeline. The figures show how easily the timeline can be embedded into a WARP's feed to notify the creation of new data objects. The index of the application maintains a reference to the latest portion of the feed so that an application can always request only the latest news.
Content is generated by social applications that store it into the butler's DB via APIs, for this purpose WARP APIs offer all the CRUD\footnote{common acronym for Create, Read, Update and Delete} functionality. Additionally, the application is allowed to compile the index file of its folder, create feeds and map DB content into network objects. Moreover, they can subscribe to be notified whenever a NOTIFY message arrives to the butler.
Social applications that run directly on the butler can be prevented from accessing to the network and distribute unencrypted data of the user, their mobile instances however could maliciously read the content storage and transfer the data somewhere else. In order to prevent this problem mobile instances can only read from the database the data that they produced. 
\section{Evaluation\label{sec:eval}}
\subsubsection{Key-Revocation}

As opposed to what was done by \cite{cachet, decent, easier, noyb} for social networking and by \cite{dynusrevoke} for cloud storage, we decided not to use proxies the key revocation process. While we do not consider the trustworthiness of proxy a real issue, since all the centralized part of such systems are generally assumed reputable, our concern are centered around scalability. Proxies are intermediaries that take part in every decrypt operation. This would imply that every data transfer in the network would at one point end in a proxy transaction, and this would include very simple operations such as, for instance, likes on Facebook. Moreover, some proxies need part of the cyphertext in order to complete the transaction which would cause an additional cost in terms of redundant network traffic that we are trying to avoid. In WARP key revocation maps into voiding cached content and re-encoding the content upon request. It is important to remember that interest for social content has generally a very short life-span. Therefore, as long as access rules are properly set when the content is generated, re-encoding might not even be used without the need to pay the cost of a proxy transaction every time the content is accessed.

Another observation that can be made about WARP's revocation procedure is whether the dimension of the FP grows linearly with the number of attributes, as discussed in section \ref{sec:keyrevoc}. Attributes are expanded only and if the content has to be re-encoded during its \textit{popular} period. Since keys are periodically changed, attributes can be shrunk again at the first key renewal. Furthermore we load-balance the problem by uniformly sub-dividing the users into buckets. As long as the number of buckets is properly chosen \cite{ballsandbins}, the chances of using high-number of attributes are low. As a ground rule we do not consider policy changes to be frequent enough to justify the network and computational overhead of using proxies. WARP pays the computational cost of re-encrypting the content when policies change, although it does that in lazy manner, namely when it is needed. Content that is not popular anymore might not have to be re-encrypted at all.

\subsubsection{Thread-Updates}

every cache should periodically ping a control channel to verify that the content of its cache is still valid, the might be a concern regarding the overhead induced by the TUs. To answer this question, the reader must consider that thread updates are network objects in the same manner any other content in the WARP network; therefore they obey to the same distribution process. In other words they can be cached and re-distributed as well. This implies that on the butler side the cost of delivering such information can be very low if the distribution chain is properly set. Moreover, on the distributor side
the cost is still fairly low considered that policy are not updated frequently and that request for updates can be sent with a granularity of choice, not to mention that they have the bandwidth occupation of only a few packets. For instance, a distributor can commit to enforce new policies only once every 30 minutes. Currently a normal CDN uses way more bandwidth only to verify which machines are still alive and which ones are not \cite{cdn}.

\subsubsection{Distribution}

Data distribution in the WARP network is intentionally kept as flexible as possible. The solutions mentioned earlier - i.e. centralized, peer-to-peer and multi-tier - are not offered by any previous work. Cachet has the advantage of offering offline data distribution which WARP does not offer because the butler has to take action after receiving a NOTIFY message, although we consider this a minor considered that the whole point of WARP is to prevent personal data to be replicated without control. Beside this aspect though, performance-wise there is not a real comparison between the two protocols. As Cachet is DHT-based content, it requires on average O(logN) steps away to reach the requester. WARP on the other hand can resolve the list of distributors in constant time without regard to the amount $N$ of users that can be in the order of tens of millions to even billions. Their use of social caching, while effective, relies on the number of connections opened and, as shown in their result, is effective only as long as requests are sent to about $40\%$ of the contacts. WARP structured approach never sends a request to more than one distributor.
\section{Discussion}

Misuse of social networks intrinsically leads to lack of privacy, and WARP cannot prevent poor decisions in content sharing. If a user is initially granted access to a content and manages to download it before the permissions have been revoked, little can be done beside trusting that the butler has not been compromised, namely the butler should cache content, for performance reasons, but not maintain a persistent copy of it. WARP, as Facebook and the other OSNs, has a best-effort approach. Once data have been delivered to a follower there is no guarantee that they will not be duplicated by the user. However, if they are not duplicated, subsequent downloads will be denied to access the content. Users can aid by maintaining a web-of-trust and avoiding sharing content with people who are known to use butlers that are not conform to the standard. 

While this problem is common to all the solutions mentioned in \ref{sec:related} WARP has the advantage of letting the user choose the right tradeoff of privacy, performance and cost.
At one extreme, a user may want the highest level of privacy and distributes his own data himself. Such user can be any artist who wants to have a fine-control over who access his material. At the other extreme, a user and his friends can organize a peer-to-peer network because they do not want to support a private company. In the first case, the cost is high but privacy is guaranteed because, not only the content is encrypted but also connections can be authenticated. In the latter case, there is no cost at all but privacy is not guaranteed if one or more users are malicious. 

In our belief, most of the users will compromise by distributing their content via a OSN, or cloud storage, because it will allow maintaining the same performance of the current network but without lacking privacy on the distribution side. 

\section{Conclusion and Future Work\label{sec:end}}

In this paper we presented WARP: a scalable ICN architecture that supports social networking with no limitations on both the user and the vendor side. The main contribution of WARP consists in offering a unique solution to the most recent privacy violations and to the current limitations of Information Centric Networking. A first version of the architecture has been implemented on top of CCNx \cite{ccnx}, working around the limitation of the NDN architecture. While it helped discovering all the issues that we tackled with this work, the implementation is far from the current design of WARP. Future work consists of re-writing the network logic of WARP and to release the first sample applications. Additionally, our priority is also to explore the scientific aspects of ICN and social network, we believe that there is a lot that can be done, especially for caching, leveraging on the information given by the social graph. Ultimately, while bulk data distribution will be most likely operated by third parties, we believe that real-time content such as status updates, tweets and messages can be distributed in a P2P fashion.
{
\bibliographystyle{abbrv}
\bibliography{biblio}
}
\end{document}